\documentclass[sigconf,nonacm]{acmart}
\AtBeginDocument{%
  }

\setcopyright{acmlicensed}
\copyrightyear{2018}
\acmYear{2018}
\acmDOI{XXXXXXX.XXXXXXX}
\acmConference[Conference acronym 'XX]{Make sure to enter the correct
  conference title from your rights confirmation email}{June 03--05,
  2018}{Woodstock, NY}
\acmISBN{978-1-4503-XXXX-X/2018/06}

\usepackage{booktabs}
\usepackage{multirow}
\usepackage{enumitem}
\usepackage{graphicx}
\usepackage{pifont}
\usepackage{xcolor}
\usepackage{afterpage}
\usepackage{placeins}
\usepackage[ruled,vlined]{algorithm2e}
\usepackage[percent]{overpic}
\usepackage{subcaption}
\usepackage[normalem]{ulem}
\usepackage{float} 

\newcommand{\sys}{EvoTune\xspace}

\usepackage{tikz}

\definecolor{stepgreen}{RGB}{39,161,91}
\definecolor{steporange}{RGB}{255,152,0}
\definecolor{stepred}{RGB}{229,20,0}
\definecolor{stepblue}{RGB}{1,172,230}

\newcommand{\cstep}[2]{%
\tikz[baseline=-0.5ex]{
  \node[
    circle,
    draw=black,
    fill=#1,
    text=white,          
    line width=0.25pt,
    inner sep=0.25pt,
    minimum size=1.05em
  ] {\bfseries\scriptsize #2};
}%
}

\begin{document}

\title{From Blind Search to Memory-Aware Evolution: Efficient DBMS Tuning via Collaborative Diagnosis and Utility-Aware Retrieval}

\author{%
Zhaoyan Hong\textsuperscript{\textdagger},
Yishen Sun\textsuperscript{\textdaggerdbl},
Xinyi Zhang\textsuperscript{\textdagger},
Zhentao Han\textsuperscript{\textdagger},
Jinhao Dong\textsuperscript{\textdagger}, \\
Wei Lu\textsuperscript{\textdagger},
Kai Xu\textsuperscript{\textdaggerdbl},
Liu Tang\textsuperscript{\textdaggerdbl},
Qi Liu\textsuperscript{\textdaggerdbl},
Xiaoyong Du\textsuperscript{\textdagger}
}

\affiliation{
  \institution{
    \textsuperscript{\textdagger}Renmin University of China,
    \textsuperscript{\textdaggerdbl}PingCAP
  }
  \country{China}
}

\email{{hongzhaoyan,xinyizhang.info,avaleph,dongjinhao,lu-wei,duyong}@ruc.edu.cn}

\email{{sunyishen,xukai,tl,liuqi}@pingcap.com}

\renewcommand{\shortauthors}{Hong et al.}

\begin{abstract}
Modern DBMSs expose multiple configurable components (e.g., knobs, query hints, and indexes) that jointly determine query performance. 
 Multi-component tuning is challenging due to the large combinatorial search space and the difficulty of learning effective tuning policies under limited feedback. 
Existing approaches still rely on blind search over the  configuration space and interaction-heavy policy learning, leading to high tuning overhead and limited performance gains. Recent advances in large language models (LLMs) enable knowledge-driven tuning, but existing LLM-based methods fail to effectively exploit online feedback and historical observations, often converging prematurely to suboptimal configurations.

In this paper, we present EvoTune, a memory-aware evolution framework for  multi-component DBMS tuning.
EvoTune first localizes a query-specific high-impact subspace via collaborative diagnosis, which combines lightweight pattern learning with LLM-based reasoning. 
It further introduces a utility-aware retrieval policy that selects informative observations based on their resulting long-term performance improvement, instead of similarity-based retrieval.
To support continual improvement, EvoTune organizes tuning feedback into a hierarchical memory and incrementally refines both subspace localization and tuning policies without requiring LLM fine-tuning. 
Extensive experiments show that EvoTune consistently outperforms state-of-the-art baselines, achieving up to 44.5\% performance improvement under the same tuning budget and reaching the best competing baseline's final performance up to 3.9$\times$ faster.
\end{abstract}


\maketitle

\section{Introduction}
Query performance is vital for modern data-driven applications and business-critical services, where even marginal increases in latency can lead to substantial economic losses~\cite{web}.
To improve query performance, a broad spectrum of configuration techniques have been extensively studied, including index selection \cite{IndexSurvey, Zhou2024Breaking, Wang2024Leveraging, Wang2025Esc, Yu2024Refactoring}, knob tuning \cite{ottertune, CDBTune, zhang_restune_2021, lambda-Tune, A-Tune-Online, onlinetune}, query rewriting \cite{LearnedRewrite, LLM-R2, R-Bot}, and plan optimization via query  hints~\cite{Bao, xu_coool_2023, Lero, woltmann_fastgres_2023}.

\begin{figure}[t]
    \centering

    \begin{subfigure}[t]{0.48\linewidth}
        \centering
        \includegraphics[width=\linewidth]{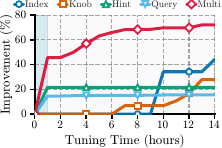}
        \caption{Single vs. Multiple Tuning.}
        \label{fig:motivate1_a}
    \end{subfigure}
    \hfill
    \begin{subfigure}[t]{0.48\linewidth}
        \centering
        \raisebox{0.6em}{\includegraphics[width=\linewidth]{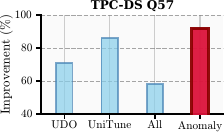}}
        \caption{Search Space Effect.}
        \label{fig:motivate1_b}
    \end{subfigure}

    \caption{\normalfont Motivating Examples.
    (a) Multi-component tuning significantly outperforms tuning single component;
    (b) Existing methods fail to localize the anomaly subspace, leading to suboptimal tuning effectiveness under the same budget. }
    \label{fig:motivate1}
\end{figure}

However, most existing approaches work separately, optimizing an individual component in isolation and are therefore insufficient to achieve globally optimal query performance. 
Although configurable components, such as knob tuning and indexing, span different layers of the query processing stack, they collectively determine the execution efficiency.
As illustrated in Figure~\ref{fig:motivate1_a}, 
tuning individual components  yields limited gains, as improvements obtained from one component are often limited and could be offset by suboptimal settings in others.
Consequently, multi-component tuning has attracted increasing attention~\cite{UDO, Unitune, ProtoX, booster}.

\begin{figure*}[t]
    \centering

    \begin{subfigure}[t]{0.24\textwidth}
        \centering
        \includegraphics[scale=1.1]{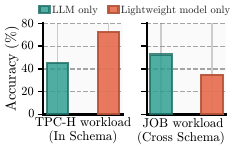}
        \caption{Model Diagnosis Accuracy.}
        \label{fig:challenge_a}
    \end{subfigure}
    \hfill
    \begin{subfigure}[t]{0.24\textwidth}
        \centering
        \includegraphics[scale=1.1]{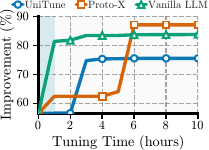}
        \caption{LLM vs. Interaction-Driven.}
        \label{fig:challenge_b}
    \end{subfigure}
    \hfill
    \begin{subfigure}[t]{0.24\textwidth}
        \centering
        \includegraphics[scale=1.2]{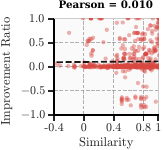}
        \caption{Similarity Retrieval Scatter.}
        \label{fig:challenge_c}
    \end{subfigure}
    \hfill
    \begin{subfigure}[t]{0.24\textwidth}
        \centering
        \includegraphics[width=0.9\linewidth]{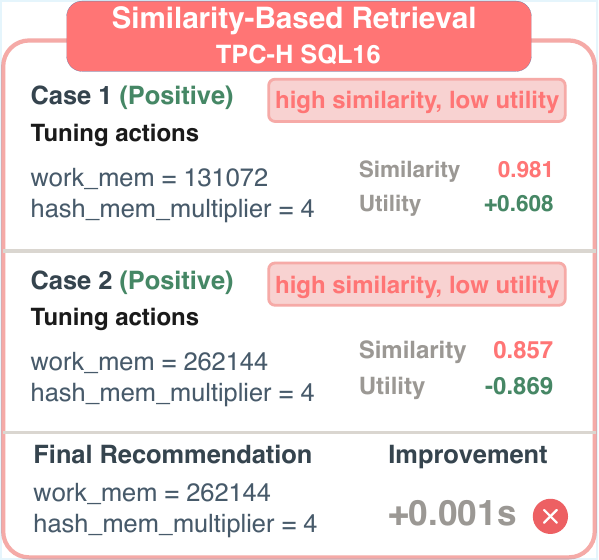}
        \caption{Similarity Retrieval Example.}
        \label{fig:challenge_d}
    \end{subfigure}

    \caption{\normalfont Preliminary Observations.
    (a) Lightweight models degrade under schema shifts, while LLMs remain stable but still unreliable across workloads;
    (b) LLM-guided tuning quickly identifies promising configurations but converges to suboptimal solutions;
    (c) Similarity is weakly correlated with improvement under similarity-based retrieval;
    (d) An example from TPC-H Q16 showing that highly similar retrieved cases may still provide limited utility for tuning.}
    \label{fig:motivate}
\end{figure*}

Despite its promise, multi-component tuning is fundamentally harder than single-component optimization for two key reasons: 
\textbf{\textit{(1) combinatorially large search space}}, as indexes, knobs, hints, and query rewrites jointly create a massive configuration space that becomes intractable to explore exhaustively under limited budgets;
\textbf{\textit{(2) hard-to-learn tuning policies}}, because the mapping  from configurations to performance is much more complex and harder to learn efficiently than for single component.
However, existing approaches~\cite{UDO, Unitune, ProtoX, booster}  do not adequately address either issue.

\textbf{\textit{First, they rely on blind search rather than adaptive localization.}}
Existing approaches are largely query-agnostic, as they do not leverage rich query information---such as SQL semantics, physical plans, and runtime statistics---to localize a query-specific anomaly subspace.
Instead, they rely on generic search strategies:
UDO~\cite{UDO} uses a hierarchical schedule that tunes indexes first and knobs later, UniTune~\cite{Unitune} selects subspaces through trial-and-error based on sparse rewards, and Proto-X~\cite{ProtoX} and Booster~\cite{booster} directly search the full joint configuration space. 
As a result, their search is either under-coordinated (e.g., UDO) or over-coordinated (e.g., Booster). 
In contrast, focusing on an anomaly subspace inferred from query information yields substantially higher efficiency under the same budget, as illustrated in Figure~\ref{fig:motivate1_b}.
\textbf{\textit{Second, they remain inefficient in learning effective tuning policies.}}
UniTune tunes configurations through ML-based online feedback optimization, which requires substantial exploration when rewards are sparse. UDO and Proto-X further depend on reinforcement-learning-based trial-and-error, requiring repeated interactions to optimize tuning decisions. As a result, these interaction-driven policy learners must accumulate many configuration-performance pairs, leading to substantial tuning overhead.
These limitations expose two fundamental challenges for multi-component tuning: (i) how to effectively localize a promising subspace for search, and (ii) how to efficiently learn high-quality tuning policies under limited feedback.

\underline{\textit{How to search effectively?}} 
Instead of searching blindly without the guidance of query information and feedback, we  propose to localize a query-specific anomaly subspace by jointly leveraging both. However, the query information is complex and implicit: existing anomaly diagnosis methods~\cite{RCRank, Dbmind} could learn execution patterns on one workload to infer anomaly-related components, but their accuracy often degrades noticeably under workload or schema shifts (detailed in Section~\ref{sec:2.1}). Moreover, tuning feedback is inherently partial: successes indicate promising space, but failures may stem from insufficient exploration rather than an incorrect subspace. \textit{This raises the challenge: how to localize a promising tuning subspace from complex query information and partial feedback, and adapt it over time as new evidence becomes available.}

\underline{\textit{How to tune efficiently?}} Recent advances in large language models (LLMs) offer a few-shot paradigm for tuning~\cite{agent_survey_1, agent_survey_2, agent_survey_3}, allowing promising actions to be inferred based on given context and reducing reliance on costly fine-tuning with large-scale interaction data. However, multi-component tuning is inherently iterative, so static few-shot inference~\cite{booster} is insufficient (detailed in Section~\ref{sec:2.1}): tuning decisions must evolve as new execution feedback accumulates. This makes the quality and organization of observations critical. Low-quality observations can be uninformative or even harmful~\cite{Noise}, while limited context windows prevent many valuable historical cases from being fully exploited. \textit{This raises the challenge of how to select, organize, and continuously refine high-quality observations so that the LLM agent can support efficient tuning under context and feedback constraints.}

In this paper, we present EvoTune, a memory-aware evolution framework that localizes query-specific  subspaces via collaborative diagnosis under partial feedback, and continuously refines tuning policies through utility-aware retrieval.
\uline{To search effectively}, EvoTune localizes the anomaly space from complex query information using two complementary diagnosticians: a lightweight data-driven model that captures recurring execution patterns, and an LLM-based Detection Agent that performs knowledge-guided reasoning for harder or out-of-distribution cases. A prior-guided bandit coordinates them, using the lightweight model as the default expert while invoking LLM reasoning and exploration when needed. To improve subsequent localization, EvoTune further refines the lightweight model with  online tuning feedback under masked supervision, avoiding overfitting to unreliable partial signals.
\underline{To tune efficiently}, EvoTune treats historical-observation retrieval as a sequential decision problem rather than static similarity matching. It learns a utility-aware retriever from tuning trajectories to select observations that are truly useful under the current tuning state, while distilling recurring successful patterns into compact experiences to alleviate context-window limitations. 
Finally, EvoTune organizes tuning observations in a memory hub and uses execution feedback to jointly refine search and retrieval throughout the tuning process. We summarize our contributions as follows.
\begin{itemize}[leftmargin=*, noitemsep, topsep=5pt]
\item We present EvoTune, an autonomous multi-component DBMS tuning framework that accelerates subspace localization and LLM-based policy learning with memory-aware evolution from execution feedback, without expensive LLM fine-tuning.

\item We propose  collaborative subspace localization via bandit-based expert selection to coordinate lightweight pattern-based prediction and LLM reasoning for reliable subspace localization.

\item We propose a learned retrieval mechanism that selects utility-aware observations and distilled experiences to support effective LLM evolution for tuning action generation.


\item Extensive experiments across workload and schema shifts show that EvoTune achieves up to 44.5\% better performance than baselines under the same tuning budget, and reaches the final performance of the best competing baseline up to 3.9$\times$ faster.
\end{itemize}
\section{Preliminary}
We first present two motivating observations in Section~\ref{sec:2.1}, then formalize the  multi-component tuning problem in Section~\ref{sec:2.2}.

\subsection{Motivating Observations}\label{sec:2.1}

In this subsection, we present two key observations that reveal fundamental gaps in effective search and efficient tuning, motivating a unified framework for robust localization and effective tuning observation utilization.

\textbf{Observation 1: Neither Lightweight Models nor LLMs Diagnose Reliably.}
Existing methods~\cite{RCRank,DBSherlock,Dbmind} rely on lightweight ML models to diagnose performance issues by learning recurring patterns from historical observations. However, such models fundamentally assume stable data distributions. As shown in Figure~\ref{fig:challenge_a}, while they achieve high accuracy on in-distribution workloads (TPC-H), their performance degrades sharply under schema shifts (JOB), revealing limited generalization.
LLMs, in contrast, perform diagnosis more like human experts by reasoning over SQL semantics, execution plans, and runtime context, enabling better generalization to out-of-distribution cases. However, such reasoning is inherently non-statistical and fails to capture recurring execution patterns, leading to inconsistent and sometimes misleading diagnoses.
These observations reveal a fundamental gap: lightweight models capture patterns but lack robustness, while LLMs generalize better but fail to exploit structured regularities. \textit{This calls for a hybrid subspace localization strategy that combines pattern-based priors with selective LLM reasoning.}

\textbf{Observation 2: Static-Similarity Retrieval Is Insufficient.} We observe a complementary trade-off between LLM-based tuning and data-driven policy learning. LLMs quickly identify promising configurations but often plateau at suboptimal solutions, while existing interaction-driven methods improve more slowly yet eventually achieve better performance (Figure~\ref{fig:challenge_b}). 
A natural approach is to combine these advantages by leveraging LLMs with feedback-driven refinement.
However, fine-tuning LLMs on the feedback data is expensive and brittle under schema or workload shifts, making in-context learning the practical alternative. 
In this setting, tuning decisions are conditioned on retrieved historical cases, which directly influence the generated actions. As a result, tuning performance becomes highly sensitive to retrieval quality.

One approach is to retrieve similar past cases based on query similarity, typically measured by cosine distance~\cite{booster}.
However, similarity is not aligned with the utility of historical observations. 
 As shown in Figures ~\ref{fig:challenge_c} and ~\ref{fig:challenge_d}, similarity is nearly uncorrelated with the final improvement, and even highly similar cases often yield negligible or negative gains.
This misalignment arises for two key reasons. First, query similarity is difficult to measure accurately for tuning, and commonly used metrics such as cosine similarity fail to capture true performance relevance. Second, tuning is inherently state-dependent: the usefulness of a historical case depends on the current tuning state, explored subspace, and past actions. Consequently, similarity-based few-shot inference remains static and fails to adapt to the evolving tuning context.
\textit{This reveals a fundamental limitation: effective tuning requires utility-aware retrieval that estimates the long-term benefit of historical cases under the current tuning state, rather than relying on static similarity.}

\subsection{Problem Formulation}\label{sec:2.2}
We aim to find an optimal configuration for a given query, particularly for slow SQL and OLAP workloads.

\noindent\textbf{Tuning Component.} 
For a given query, the tuning search space consists of four classes of actions:
(1) \textit{query rewriting}, which restructures SQL statements into semantically equivalent forms but with higher performance; (2) \textit{query-level knob tuning}, which adjusts execution-related DBMS parameters that affect query behavior and can be set on a per-query basis (e.g., \textit{work\_mem}); (3) \textit{optimizer hints}, which inject explicit directives to guide the query optimizer to generate better query plans  (e.g., \textit{enable\_seqscan}); and (4) \textit{index selection}, which  recommends indexes to accelerate query processing. 




\noindent\textbf{Multi-Component Tuning.}
Suppose there are $m$ tunable components, each component corresponds to a configuration subspace $\Theta_i$ $(1 \le i \le m)$, and a configuration setting is denoted by $\theta_i \in \Theta_i$.
The full configuration space is defined as the Cartesian product $\Theta = \Theta_1 \times \Theta_2 \times \cdots \times \Theta_m$.
Given a target query $q$ and tuning objective $f$, the multi-component tuning problem aims to find a configuration:
\begin{equation}
\arg\min_{(\theta_1, \ldots, \theta_m) \in \Theta}
f(\theta_1, \ldots, \theta_m; q),
\end{equation}
where $f(\cdot ; q)$ measures the execution performance of query 
$q$ under a given configuration, typically in terms of query execution time.


\noindent\textbf{Two-Stage Multi-Component Tuning.}
Given a query $q$, the goal is to efficiently optimize performance within a high-dimensional configuration space. Instead of directly searching the full space $\Theta$, we formulate the problem as a two-stage decomposition:

\textit{(1) Tuning Space Localization.}
The goal is to determine a subset of bottleneck-related components $\mathcal{B}$, i.e., components whose configuration adjustments can lead to measurable performance improvements for the target query. Formally, we define:
\begin{equation}
\mathcal{B} = \{ b_1, b_2, \ldots, b_n \} \subseteq \{1, 2, \ldots, m\}, \qquad 0 < n \le m,
\end{equation}
which induces a query-specific subspace
$\Theta_{\mathcal{B}} = \prod_{i \in \mathcal{B}} \Theta_i$. Note that the subspace $\Theta_{\mathcal{B}}$ is not fixed and can be iteratively refined as new evidence becomes available.

\textit{(2) Tuning over Subspace.}
Given the localized  subspace $\Theta_{\mathcal{B}}$, the objective is to optimize the configuration within this reduced space:
\begin{equation}
\arg\min_{(\theta_{b_1}, \ldots, \theta_{b_n}) \in \Theta_{\mathcal{B}}}
f(\theta_{b_1}, \ldots, \theta_{b_n}; q).
\end{equation}



\begin{figure*}[ht]
    \centering 
    \includegraphics[width=1\textwidth]{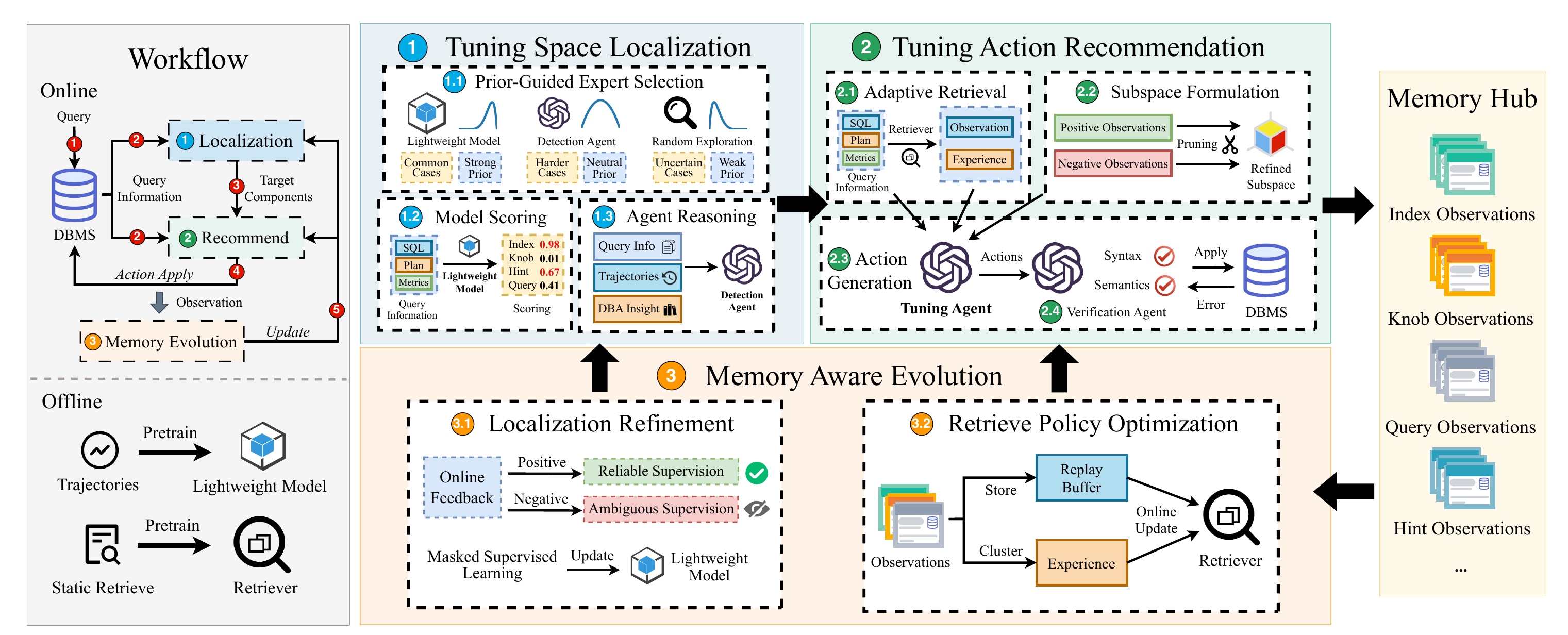}
    \caption{System Overview of EvoTune.} \label{overview} 
\end{figure*}

\section{Overview}

Figure~\ref{overview} presents the end-to-end workflow of \sys. It consists of three tightly coupled modules: \cstep{stepblue}{1} tuning space localization, \cstep{stepgreen}{2} tuning action recommendation, and \cstep{steporange}{3} memory-aware evolution. 
These modules are instantiated as sequential stages in an online feedback-driven loop, where observations continuously improve subsequent localization and recommendation.

\noindent\underline{Workflow.} 
In the \textbf{offline} phase, \sys pretrains a lightweight detection model and a learned retriever to warm-start localization and recommendation modules.
In the \textbf{online} phase, for each target query, \cstep{stepred}{1} it first collects \textit{query information}, including the SQL statement, execution plan, and runtime metrics. \cstep{stepred}{2} This information is then passed to both the localization and recommendation stages. \cstep{stepred}{3} The localization stage identifies the target tuning components, determining where optimization should focus. \cstep{stepred}{4} Conditioned on the localized subspace, the recommendation stage retrieves relevant observations and experiences to generate and apply tuning actions. \cstep{stepred}{5} The resulting outcome is written to a memory hub as a new observation, based on which the memory-aware evolution module updates both the localization and recommendation strategies.
Tuning terminates when  the query performance reaches a target threshold or when no improvement is observed over several consecutive iterations.
Next, we describe the modules in detail.

\cstep{stepblue}{1} \textbf{Tuning Space Localization.}
Given the collected \textit{query information}, \sys identifies bottleneck-related tuning components $\mathcal{B}$ through \cstep{stepblue}{1.1} a prior-guided expert selection mechanism that coordinates three localization choices: \cstep{stepblue}{1.2} component scoring with the lightweight model, \cstep{stepblue}{1.3} agent reasoning with the Detection Agent, and random exploration. This mechanism balances efficiency, reasoning flexibility, and exploration, allowing \sys to rely on lightweight prediction for common cases, invoke agent reasoning for cases that lightweight model fails to solve, and retain exploration when both become unreliable. The lightweight model takes the \textit{query information} as input and estimates the likelihood that different tuning components exhibit performance bottlenecks; it is continuously refined by \cstep{steporange}{3.1} using feedback from tuning outcomes. Guided by DBA insight and trajectories, i.e., previous tuning observations, the Detection Agent performs structured reasoning over the \textit{query information} to infer the target tuning components.

\cstep{stepgreen}{2} \textbf{Tuning Action Recommendation.}
Given the target tuning components, \sys recommends actions via a structured workflow. 
\cstep{stepgreen}{2.1} \emph{Adaptive Retrieval}: 
Retrieve historical observations with their experiences that are expected to be informative via a learned strategy (\cstep{steporange}{3.2})  based on the target components and query information; 
\cstep{stepgreen}{2.2} \emph{Subspace Formulation}: Construct a refined subspace by pruning unpromising configurations to constrain search within a meaningful optimization region; 
\cstep{stepgreen}{2.3} \emph{Action Generation}: Generate tuning actions within the refined subspace by reflective reasoning over the retrieved history; 
\cstep{stepgreen}{2.4} \emph{Action Evaluation}: Validate the syntactic and semantic correctness of actions through interaction with a Verification Agent, and apply validated actions to evaluate query performance and update the memory hub with the observations.

\cstep{steporange}{3} \textbf{Memory-Aware Evolution.}
To enable continuous adaptation, \sys introduces a memory-aware evolution mechanism that leverages tuning feedback to jointly optimize subspace localization and action recommendation modules through lightweight parameter updates, without fine-tuning the underlying LLMs.

\cstep{steporange}{3.1} \textit{\underline{Feedback-Driven Localization Refinement.}}
To improve the subspace localization over time, \sys refines the lightweight model using feedback from tuning outcomes. Learning from online tuning feedback is challenging due to asymmetric and incomplete supervision: positive feedback provides reliable evidence for validated tuning components, whereas negative feedback is often ambiguous and may arise from insufficient exploration rather than incorrect localization. To address this issue, \sys applies masked supervised learning to ambiguous negative feedback, thereby avoiding incorrect hard supervision and enabling more reliable localization refinement over time (as detailed in Section~\ref{sec:4.1.2}).

\cstep{steporange}{3.2} \textit{\underline{Retrieve Policy Optimization.}} 
The effectiveness of tuning actions depends critically on the historical observations provided to the Tuning Agent, which reuses past feedback to guide recommendations. 
To make historical reuse adaptive under limited context windows, \sys clusters observations into distilled experiences (Section~\ref{sec:5.1}) and learns a utility-aware retriever (Section~\ref{sec:5.2}) to select informative observations and experiences for recommendation. 
Given the inherently multi-iteration nature of database tuning~\cite{CDBTune}, \sys models the tuning process as a sequential decision problem and optimizes the retriever to maximize long-term utility under current state. 
As new observations are collected, they are stored in a replay buffer for retriever updates and clustered periodically into experiences for future recommendations.
\section{Collaborative Detection}
This section presents the collaborative localization mechanism, which combines a lightweight data-driven model (Section~\ref{sec:4.1}) and an LLM-based Detection Agent (Section~\ref{sec:4.2}) via an expert-based  coordination (Section~\ref{sec:4.3}) for promising space localization.

\subsection{Data-Driven Component Scoring}\label{sec:4.1}
We first describe the construction of the  lightweight detection model, followed by its localization refinement mechanism.

\subsubsection{Component Scoring Model}\label{sec:4.1.1}
We formulate tuning-component localization as a multi-label classification task  over a predefined component set $\mathcal{C}=\{1,\dots,m\}$.
Given the \textit{query information} as input, the detection model predicts  predicts component-level scores indicating the benefit of tuning each component.
Formally, the detection model outputs component scores $\hat{\mathcal{Y}}=\{\hat{\mathcal{Y}}_1,\ldots, \hat{\mathcal{Y}}_m\}$, where $\hat{\mathcal{Y}}_b\in[0,1]$ estimates the probability that the query  performance can be improved by tuning the component $b$. 
The diagnosed component set $\mathcal{B}$ is obtained by thresholding $\hat{\mathcal{Y}}$ at $0.5$, and could contain multiple components for compound anomalies.

\noindent\textbf{Query Featurization.}
Given a target query, \sys collects its \textit{query information} and represents it as a four-tuple $\mathcal{Q}=(S,P,I,E)$, capturing complementary aspects of query semantics, execution behavior, and system dynamics:

\begin{itemize}[leftmargin=0.3cm]
\item \emph{\underline{SQL Statement ($S$)}} 
is treated as a textual sequence, where each token 
(e.g., \texttt{SELECT}) captures query intent as well as table or attribute references.

\item \emph{\underline{Execution Plan ($P$)}} 
is represented as a structured execution trace composed of physical operators
(e.g., sequential scan and join) together with their associated cost statistics and dependencies.

\item \emph{\underline{Internal Metrics ($I$)}} record DBMS statistics collected during query execution. 
These metrics reflect query behaviors such as row access, block I/O, and buffer miss.

\item \emph{\underline{External Metrics ($E$)}} are  multivariate time series tracking system-level indicators (e.g., CPU utilization, memory load, I/O throughput). 
As query execution durations vary, 
we apply linear interpolation to align all time series to a uniform length.
\end{itemize}
We encode each modality using a dedicated encoder~\cite{RCRank}: BERT~\cite{BERT} for query statements, QueryFormer~\cite{QueryFormer} for execution plans,  a multi-layer perceptron (MLP) for internal  metrics, and a 2D-CNN model for external metrics. 
The resulting representations are fused via a cross-modal Transformer, followed by an MLP that outputs component-level logits.

\noindent\textbf{Impact Label Definition.}
To supervise component scoring, we derive an outcome label indicating whether the selected components are worth tuning.
Given a diagnosed component set  $\mathcal{B}$,  we  define a binary tuning outcome after applying action $\mathcal{A}$ over $\mathcal{B}$:
\begin{equation}\label{eq:outcome}
\mathcal{Y}
=
\mathbb{I}\!\left[
\frac{\mathcal{T}(\mathcal{Q})-\mathcal{T}(\mathcal{Q},\mathcal{A})}
{\mathcal{T}(\mathcal{Q})}
\ge \delta
\right].
\end{equation}
Based on this outcome, the component-level impact label for each component $b\in\mathcal{C}$ is defined as:
\begin{equation}\label{eq:root}
\mathcal{Y}_b
=
\mathbb{I}[b\in\mathcal{B}] \cdot \mathcal{Y}.
\end{equation}
Here, $\mathcal{T}(\mathcal{Q})$ and $\mathcal{T}(\mathcal{Q},\mathcal{A})$ denote the query latency before and after applying action $\mathcal{A}$, respectively.
$\delta$ is a minimum improvement threshold (set to $5\%$) to filter noise, and $\mathbb{I}[\cdot]$ denotes  indicator function.
Thus, components in $\mathcal{B}$ are labeled positive only if the action yields sufficient improvement.

\subsubsection{Continual Refinement via Masked Supervision} \label{sec:4.1.2}

A key advantage of the detection model is its ability to capture fine-grained patterns from data via lightweight parameter updates, compared with LLM agents.
However, online tuning feedback is inherently asymmetric and incomplete for subspace localization.
A positive outcome provides reliable evidence that tuning component $b$ is beneficial for query $i$.
In contrast, a negative outcome is often ambiguous: it may indicate that $b$ is truly ineffective, but it may also simply result from insufficient exploration over that component.
Thus, directly applying a standard binary cross-entropy (BCE) loss to all negative labels is inappropriate, because it would treat under-explored components as true negatives and discourages further exploration, introducing false-negative bias and suppressing potentially beneficial components.

To address this issue, we adopt a masked binary cross-entropy objective that retains only reliable supervision.
Let $n_{i,b}$ denote the number of tuning trials  on  component $b$ for query $i$, and let $\eta$ be a minimum exploration threshold.
A negative label is regarded reliable only when $n_{i,b}\geq \eta$.
Besides, components not selected for tuning are treated as unlabeled.
Accordingly, we define the supervision mask as:
\begin{equation}
m_{i,b} =
\begin{cases}
1, & \mathcal{Y}_{i,b}=1,\\
1, & \mathcal{Y}_{i,b}=0 \ \land\ n_{i,b}\geq \eta,\\
0, & \mathcal{Y}_{i,b}=0 \ \land\ n_{i,b}<\eta,\\
0, & b \notin \mathcal{B}.
\end{cases}
\end{equation}
We minimize the following masked BCE loss:
\begin{equation}
\scalebox{0.91}{$
\mathcal{L}_{b}
=
-\frac{1}{\sum\limits_{i}\sum\limits_{b=1}^{m} m_{i,b}}
\sum\limits_{i}\sum\limits_{b=1}^{m}
m_{i,b}
\left[
\mathcal{Y}_{i,b}\log \hat{\mathcal{Y}}_{i,b}
+
(1-\mathcal{Y}_{i,b})\log\!\bigl(1-\hat{\mathcal{Y}}_{i,b}\bigr)
\right]
$}
\end{equation}
In this way, \sys always learns from verified positive signals and sufficiently explored negative signals, while ignoring unreliable negatives caused by limited  exploration.

\subsection{Knowledge-Driven Reasoning}\label{sec:4.2}

The Detection Agent infers tuning components with domain knowledge and historical trajectories to ensure robust diagnosis.

\noindent\textbf{Domain Knowledge-Guided Reasoning.}
We encode DBA insights as structured prompts to emulate DBA-style reasoning for bottleneck validation. Representative insights include:

\begin{itemize}[leftmargin=*]

    \item \emph{\underline{Plan Suboptimality.}}
    A plan bottleneck typically arises from suboptimal access paths, join orders, or join strategies that introduce unnecessary overhead, such as redundant sorting, full table scans, or excessive intermediate results.
  Typical evidence includes cases where the optimizer prefers sequential scans despite the presence of indexes under high selectivity, incurs extra sorting due to unordered access paths, exhibits hash join spill-to-disk behavior.

      \item \emph{\underline{Index Bottleneck.}} Index bottlenecks are characterized by large-range scans on selective predicates. Typical evidence includes 
    a large gap between scanned and returned rows, 
    predicates that cannot be effectively exploited by indexes, 
    and repeated heap/table access due to non-covering indexes. 

    \item \emph{\underline{Knob Misconfiguration.}} Knob tuning becomes important when analytical operators spill to disk or lack sufficient memory or parallel resources. Typical evidence includes disk-backed temporary tables, 
    underutilized parallel workers, 
    and high-cost join or aggregation execution under limited resources. 

    \item \emph{\underline{Query Formulation Issues.}}
    Query rewriting is prioritized when the SQL structure itself introduces unnecessary work. Typical evidence includes 
    implicit type conversions (blocking B-Tree access and triggering full scans),
    overlong \texttt{IN} lists.
\end{itemize}

\noindent\textbf{Trajectory-Guided Reasoning.}
For each query $i$, we leverage its tuning trajectory to refine component confidence and avoid repeatedly exploring unpromising directions.
Specifically, for each component $b$, \sys maintains the numbers of successful and failed tuning attempts, denoted by $s_{i,b}$ and $f_{i,b}$. 
These statistics are incorporated into structured prompts to guide the Detection Agent according to the following principles:
\begin{itemize}[leftmargin=0.3cm]
    \item If $s_{i,b} > 0$ and $f_{i,b} = 0$, prioritize component $b$, since past interventions consistently validate it as an actionable bottleneck.
    \item If $s_{i,b} = 0$ and $f_{i,b} > 0$, deprioritize component $b$ and encourage exploration of alternative components, since previous interventions on $b$ have not led to improvement.
    \item If $s_{i,b} > 0$ and $f_{i,b} > 0$, keep component $b$ under consideration and let the Detection Agent make a more nuanced judgment based on DBA knowledge and the available evidence.
\end{itemize}


Based on the above knowledge and trajectory signals, we construct a structured prompt and populate it with database context and selected query information  (listed in Appendix~A.1~\cite{appendix}), based on which the Detection Agent infers the tuning components $\mathcal{B}$.

\subsection{Prior-Guided Expert Selection}\label{sec:4.3}
As discussed in Section~\ref{sec:2.1}, the two diagnosis paradigms exhibit complementary strengths for tuning-space localization:
the \emph{lightweight detection model} efficiently captures recurring patterns from historical data, while the \emph{LLM-based Detection Agent} reasons more flexibly over query evidence and is thus better suited for harder or out-of-distribution cases when lightweight model fails. 

To address this uncertainty, we adaptively select the most suitable diagnostician for each query by formulating diagnostician selection as a  multi-armed bandit problem~\cite{bandit}, where each expert is treated as an arm and selected at each tuning step.
To encode prior knowledge about the relative strengths of different experts, we incorporate asymmetric priors, guiding selection using both prior knowledge and downstream tuning feedback.

\noindent\textbf{Expert Arms.}
We define three expert arms $E=\{e_{\mathrm{lm}},\, e_{\mathrm{llm}},\, e_{\mathrm{rand}}\}$,
where $e_{\mathrm{lm}}$, $e_{\mathrm{llm}}$, and $e_{\mathrm{rand}}$ denote the lightweight model, the Detection Agent, and random exploration, respectively.
The random arm serves as a fallback to explore alternative subspaces when both diagnosticians are uncertain, preventing overcommitment to potentially suboptimal regions.

\noindent\textbf{Prior-Guided  Expert Selection.}
For each arm $e \in E$, \sys maintains a Beta distribution $\theta_e \sim \mathrm{Beta}(\alpha_e,\beta_e)$ to model its expected utility for tuning-space localization, where $\alpha_e$ and $\beta_e$ correspond to the accumulated counts of positive and negative feedback.
\sys adopts Thompson sampling for expert selection: at each tuning iteration $t$, it draws $\tilde{\theta} \sim \mathrm{Beta}(\alpha_e,\beta_e)$ for each arm and selects $ e_t = \arg\max_{e \in E} \tilde{\theta}_e$.
To encode prior knowledge about the relative strengths of different diagnosticians, the initialization is deliberately asymmetric: the lightweight model is assigned the strongest prior as it achieves higher accuracy on in-distribution workloads, the Detection Agent receives a weaker but still competitive prior for harder or out-of-distribution cases, and random exploration receives the weakest prior.
Accordingly, the initialization satisfies $\mathbb{E}[\theta_{e_{\mathrm{lm}}}] >
\mathbb{E}[\theta_{e_{\mathrm{llm}}}] >
\mathbb{E}[\theta_{e_{\mathrm{rand}}}]$, which is achieved by using Beta distributions with different parameterizations.
In practice, we use $\mathrm{Beta}(8,2)$, $\mathrm{Beta}(2,2)$, and $\mathrm{Beta}(2,8)$ for the three arms, respectively.

\noindent\textbf{Feedback-Driven Bandit Update.}
After selecting expert $e_t$, \sys optimizes within the localized subspace $\mathcal{B}_t$ for $\eta$ iterations, rather than immediately performing another expert selection, as a promising subspace may require multiple iterations to reveal its utility and would otherwise be prematurely discarded.
Then, \sys summarizes the downstream outcome of $\mathcal{B}_t$ into a binary impact label $\mathcal{Y}_t$ (Eq.~\ref{eq:outcome}).
The posterior parameters of the selected arm are then updated as: 
\begin{equation}
\alpha_{e_t} \leftarrow \alpha_{e_t}+\mathcal{Y}_t, \qquad
\beta_{e_t} \leftarrow \beta_{e_t}+(1-\mathcal{Y}_t).
\end{equation}
Thus, experts that repeatedly yield effective localization decisions receive stronger posterior support, while those leading to weak localization outcomes are gradually down-weighted.
To ensure query-specific adaptation, the Beta distributions are re-initialized for each query.

\section{Tuning Action Recommendation}

The tuning process can be naturally formulated as a Markov Decision Process (MDP), since tuning is inherently a sequential decision-making problem in which each action affects both the immediate reward and the subsequent tuning state~\cite{CDBTune}.
When leveraging LLM agents for tuning, particularly in retrieval-augmented settings, we argue that tuning decisions are conditioned not only on the current state but also on retrieved historical cases, which encode past tuning experiences and provide essential context for decision making.
Formally, at iteration $t$, we define the current state as $\mathcal{S}_t=(\mathcal{Q}_t,\mathcal{B}_t)$, where $\mathcal{Q}_t$ denotes the current query information and $\mathcal{B}_t$ the localized tuning components. The memory hub $\mathcal{M}_t={\mathcal{O}_1,\dots,\mathcal{O}_N}$ stores historical tuning observations accumulated up to iteration $t$.
For LLM agents, a retrieval policy $\mu(\mathcal{O}\mid \mathcal{S}_t,{M}_t)$ selects relevant past observations, based on which the LLM generates actions following $p_{\text{LLM}}(\mathcal{A}_t\mid \mathcal{S}_t,\mathcal{O})$. The overall tuning policy is thus given by:
\begin{equation}
\pi(\mathcal{A}\mid \mathcal{S}_t,M_t)=\sum_{\mathcal{O}\in M_t}\mu(\mathcal{O}\mid \mathcal{S}_t,M_t)\,p_{\text{LLM}}(\mathcal{A}\mid \mathcal{S}_t,\mathcal{O}).
\end{equation}
To improve the tuning policy, $p_{\text{LLM}}$ is instantiated with an off-the-shelf LLM, enabling seamless upgrades as more advanced models become available. Rather than relying on LLM fine-tuning, which is costly and prone to overfitting under distribution shifts~\cite{DBLP:journals/bdcc/WuCLWLLHHPMHHGCFLTNW25}, we focus on improving decision quality through external memory and retrieval.
This highlights the central role of memory construction and retrieval in tuning. Accordingly, \sys first organizes historical feedback into a hierarchical memory hub (Section~\ref{sec:5.1}), then learns an adaptive retrieval policy over this memory (Section~\ref{sec:5.2}), and finally generates and validates executable actions (Section~\ref{sec:5.3}).

\subsection{Experience Distillation and Evolution}\label{sec:5.1}
To support the sequential tuning process with reusable memory, \sys organizes historical feedback into a hierarchical memory hub with two levels: \emph{instance-level observations} and \emph{cluster-level experiences}. The former preserve fine-grained evidence from individual tuning attempts, while the latter capture stable action patterns shared across similar observations. This two-level design enables effective reuse by combining query-relevant evidence with higher-level, transferable tuning guidance.

\noindent\textbf{Observation Construction.}
After each tuning iteration, \sys records the tuning result as an instance-level observation
\begin{equation}
\mathcal{O}=(\mathcal{Q}, \mathcal{B}, \mathcal{A}, \mathcal{T}, \mathcal{Y}, \mathcal{F}),
\end{equation}
where $\mathcal{Q}$, $\mathcal{B}$, $\mathcal{A}$, and $\mathcal{T}$ denote the query information, localized tuning components, applied action, and observed execution performance, respectively, and $\mathcal{Y}\in{0,1}$ denotes the outcome label defined in Eq.~\ref{eq:outcome}.
To improve transferability while keeping historical feedback grounded, \sys further derives a reflection $\mathcal{F}$ from the current tuning result by abstracting the query and execution characteristics, the applied action, and the resulting outcome.
According to  label $\mathcal{Y}$, the reflection $\mathcal{F}$ is categorized into two forms:
\begin{itemize}[leftmargin=*]
    \item \emph{Positive reflection}: captures why the applied action is effective  under the observed query and execution evidence. 
    \item \emph{Negative reflection}: abstracts why the applied action is ineffective under the observed query and execution evidence. 
\end{itemize}

\noindent\textbf{Experience Distillation.}
While reflections capture fine-grained feedback for individual observations, they remain tied to specific cases. EvoTune therefore further abstracts observations into experiences for more transferable reuse. To avoid noisy or overly case-specific summaries, it first clusters similar observations and then distills each cluster into an experience that captures a stable tuning pattern, including representative  execution characteristics, recurring actions, and their overall effects.
For example, \sys distills the following experiences from historical observations:
\begin{itemize}[leftmargin=*]
    \item \emph{Positive experience}: when execution remains scan-dominated and the number of planned/launched workers has already reached the limit of \texttt{max\allowbreak\_parallel\allowbreak\_workers\allowbreak\_per\allowbreak\_gather}---the PostgreSQL parameter that caps the maximum number of parallel workers used by a single gather node---increasing this limit may help by enabling more parallel scan workers.

    \item \emph{Negative experience}: when the dominant cost comes from repeated large-scale aggregations and rescans, and the relevant sort/hash operators already fit in memory without spilling to disk, simply increasing \texttt{work\_mem}---the per-operator memory budget for in-memory sorting and hashing---may fail to improve performance, as the bottleneck is not memory-bound but stems from the data-intensive execution pattern.
\end{itemize}



To support clustering, \sys represents each observation using a joint embedding constructed by concatenating the query information embedding (Section~\ref{sec:4.1.1}), the action embedding produced by a BERT encoder, and a scalar latency-improvement feature.
Considering that online tuning observations are inherently heterogeneous and noisy, and the number of underlying latent tuning patterns is unknown in advance, \sys adopts HDBSCAN~\cite{HDBSCAN} for observation clustering. 
HDBSCAN can identify clusters of varying densities without requiring a predefined number of clusters, while leaving noisy observations unassigned rather than forcing them into unreliable clusters. For each cluster, \sys invokes a Summary Agent that summarizes the reflections derived from its member observations into a single cluster-level experience.

\noindent\textbf{Experience Evolution.}
As tuning observations arrive continuously during online tuning, \sys incrementally maintains the experience instead of reclustering all observations from scratch. For each cluster, \sys keeps a prototype, defined as the average embedding of its member observations, and a radius threshold, defined as the distance from the prototype to the farthest member observation. Given a new observation, \sys computes its distance to all cluster prototypes and assigns it to the nearest cluster if the distance falls within that cluster’s radius. In that case, the observation is merged into the cluster. Otherwise, it is stored in a \emph{pending buffer}. When the buffer accumulates sufficient observations, \sys reclusters them to identify emerging tuning patterns that are not captured by existing clusters.

\subsection{Adaptive Memory Retrieval}\label{sec:5.2}
Given the memory hub, the central challenge is to retrieve historical observations that are most beneficial for the tuning process. Due to the sequential nature of tuning as an MDP, the utility of a retrieved observation cannot be determined in isolation, but depends on its long-term contribution to tuning performance. 
The observations to be retrieved should evolve as the tuning proceeds to support continual adaptation of the agent’s tuning policy $p_{\text{LLM}}(\mathcal{A}_t\mid \mathcal{S}_t,\mathcal{O})$, rendering static retrieval inadequate.
Therefore, we propose a reinforcement learning formulation to optimize retrieval decisions based on their long-term impact on tuning performance.

To this end, \sys leverages tuning feedback to train a parameterized retriever that learns a Q-function over current state and historical observations, which guides retrieval by estimating their expected utility for long-term gain.
Importantly, the current state is not limited to the SQL statement, but also includes execution plans and runtime metrics, which are continuously shaped by previously applied actions and observed feedback.
The retriever is optimized via soft Q-learning to prioritize observations that maximize long-term utility for iterative tuning.
We define the reward as the immediate reduction in query latency after applying the action induced by the retrieved observation:
\begin{equation}
\scalebox{0.9}{$
\mathcal{R}_t = \mathcal{T}(\mathcal{S}_t) - \mathcal{T}(\mathcal{S}_{t+1}), \;
\mathcal{A}_t \sim p_{\text{LLM}}(\cdot \mid \mathcal{S}_t, \mathcal{O}_t), \;
\mathcal{S}_{t+1} = f(\mathcal{S}_t, \mathcal{A}_t)$,}
\end{equation}
where $\mathcal{O}_t$ denotes the retrieved observation at iteration $t$.
The Q-value captures the expected cumulative performance improvement when selecting an observation under the current state.
Specifically, for each candidate historical observation, the learned retriever takes as input a feature vector formed by concatenating: (i) the query information embedding of $\mathcal{Q}_t$, (ii) a binary encoding of the diagnosed components $\mathcal{B}_t$, and (iii) the embedding of the candidate observation.
The retriever is implemented as a two-layer MLP that outputs a scalar Q-value.

Inspired by \cite{memento},  we  adopt a maximum-entropy reinforcement learning objective~\cite{haarnoja2018soft} to optimize the retrieval policy while encouraging diversity among retrieved observations.
\begin{equation}
\label{eq:objective}
J(\mu)=\mathbb E_{p_\mu}\left[\sum_{t=0}^{T-1}\left(\mathcal{R}_t+ \mathcal H(\mu(\cdot\mid \mathcal{S}_t, M_t))\right)\right],
\end{equation}
where $\mathcal{H}$ denotes the entropy.
Under the maximum-entropy objective in Eq.~\ref{eq:objective}, the optimal retrieval policy induced by the soft Q-function takes a Boltzmann
form over candidate observations:
\begin{equation}
\mu^*(\mathcal{O} \mid \mathcal{S}_t, M_t)
=
\frac{\exp\big(Q^*(\mathcal{S}_t, M_t, \mathcal{O})\big)}
{\sum_{\mathcal{O}' \in M} \exp\big(Q^*(\mathcal{S}_t, M_t, \mathcal{O}')\big)},
\end{equation}
it assigns larger retrieval probabilities to observations with higher Q-values.
Using tuning trajectories, the retriever is trained by minimizing temporal-difference (TD) loss. 
Specifically, given a transition $(\mathcal{S}_t, M_t,$ $\mathcal{O}_t, \mathcal{R}_t,$ $ \mathcal{S}_{t+1}, M_{t+1})$, 
the state value under the optimal retrieval policy is defined as:
\begin{equation}
V^*(\mathcal{S}_t, M_t)
=
\log
\sum_{\mathcal{O}' \in M_t}
\exp\big(Q^*(\mathcal{S}, M_t, \mathcal{O}')\big),
\end{equation}
which measures the maximum long-term utility achievable at retrieval state $(\mathcal{S}_t, M_t)$ when candidate observations are selected according to the optimal retrieval policy.
The TD target is defined using a soft Bellman backup:
$Q^\text{target}_t
= \mathcal{R}_t + \gamma V(\mathcal{S}_{t+1}, M_{t+1}).$
Here, $\gamma \in [0,1)$ is the discount factor for future retrieval utility.
The Q-network is optimized by minimizing the squared TD error:
\begin{equation}\label{eqa:dt}
\mathcal{L}_{\text{TD}}(\theta)
= \mathbb{E}
\Big[
\big(
Q_\theta(\mathcal{S}_t, M_t, \mathcal{O}_t) - Q^\text{target}_t
\big)^2
\Big].
\end{equation}

\noindent\textbf{Offline Pretraining.} Learning a promising retrieval policy from a randomly initialized network is challenging, as in the initial training stage the retriever tends to select uninformative observations, leading to sparse reward signals and hindering stable and efficient convergence.
To mitigate this issue, we adopt an offline training strategy by first constructing trajectories using a similarity-guided bootstrap policy. 
 Specifically, with probability $1-\epsilon$, the bootstrap policy selects an observation from the top-$k$ most similar candidates, and with probability $\epsilon$, it samples an observation uniformly at random from the remaining memory. Although similarity is insufficient as the final retrieval criterion, it serves as a useful weak prior for guiding early exploration, while random exploration improves coverage over historical observations and prevents the retriever from collapsing to a purely similarity-based policy.
\noindent\textbf{Online Continual  Learning.} 
During online retrieval, the retriever selects the top-$k$ observations with the highest Q-values, together with the associated cluster-level experiences  to construct a compact yet informative context for the Tuning Agent.
Newly collected tuning trajectories are continuously stored in a replay buffer. Once the buffer reaches 256 samples, the Q-network is updated using randomly sampled batches to further minimize the TD loss in Eq.~\ref{eqa:dt}, enabling continual refinement of the retrieval policy under evolving query contexts.


\subsection{Action Generation and Evaluation}  \label{sec:5.3}
Given the retrieved observations $\mathbb{O} = \{\mathcal{O}_1, \ldots, \mathcal{O}_k\}$, \sys follows a structured workflow to produce tuning actions: it constructs a refined tuning subspace to guide recommendation in promising regions, and then generates candidate actions with reflective reasoning and validates them through execution-oriented evaluation.


\noindent\textbf{Subspace Formulation.}
Before action generation, \sys first defines a \emph{basic tuning space} from the localized tuning components $\mathcal{B}$: $\Theta_{\mathcal{B}} = \prod_{b \in \mathcal{B}} \Theta_b,$ which provides the Tuning Agent with the allowable operation space for each identified component.
Based on this basic space, \sys further leverages the retrieved observations to refine it into a \emph{promising subspace} $\Theta^*$ by pruning ineffective choices and narrowing the search toward historically beneficial regions.
Specifically, for discrete spaces such as indexes and query rewrites, \sys prunes candidates with repeatedly ineffective outcomes in retrieved observations; for continuous or configurable spaces such as knobs and hints, it narrows the admissible values to regions that previously achieved higher utility.
Importantly, $\Theta^*$ serves as a guidance prior rather than a hard constraint, allowing the Tuning Agent to deviate from it when necessary.
Details are deferred to Appendix~A.2~\cite{appendix}.


\noindent\textbf{Generation with Reflective Reasoning.}
Conditioned on $\mathcal{Q}$, $\mathcal{B}$, $\mathbb{O}$, and $\Theta^*$, the Tuning Agent generates tuning actions via \emph{reflective reasoning}~\cite{Reflexion}. Specifically, at each iteration $t$, \sys prompts the agent to reflect on the outcomes of previous attempts, analyze why past actions in retrieved observations succeeded or failed, and produce the next evidence-supported action:
$
\mathcal{A}_{t+1} \sim p_{\text{LLM}}\bigl(\mathcal{Q}_t, \mathcal{B}_t, \mathbb{O}_t, \Theta^*_t\bigr).
$
To ensure executability, \sys requires the generated action to follow predefined templates for each tuning component, together with DBMS-specific syntax constraints, so that it can be directly applied to the target system.

\noindent\textbf{Action Evaluation Pipeline.}
\sys uses the Verification Agent to validate each generated action for syntactic validity and execution feasibility. Specifically, the agent checks whether the action conforms to the DBMS-specific syntax and whether it can be successfully applied or executed on the target system. Invalid actions are retried within a fixed attempt budget, and persistent failures are recorded in the memory hub to avoid repeated invalid attempts in future iterations.
For query rewriting, \sys further verifies semantic equivalence with the original SQL through a hierarchical, tool-first pipeline. It first applies QED~\cite{QED}, a semantic equivalence checker for rewritten SQL. If QED is inconclusive, \sys executes both queries and compares their results. If a mismatch remains after up to $n$ rounds of iterative repair, \sys falls back to the original SQL.
After execution, \sys assigns the outcome label $\mathcal{Y}$ and generates the reflection $\mathcal{F}$, then stores the resulting observation $(\mathcal{Q}, \mathcal{B}, \mathcal{A}, \mathcal{T}, \mathcal{Y}, \mathcal{F})$ in the memory hub.
\section{Evaluation}
We detail the experimental setup in Section~\ref{sec:6.1} and present the end-to-end evaluation  in Section~\ref{sec:6.2}. We then validate the design of \sys, including space localization in Section~\ref{sec:6.3}, action recommendation in Section~\ref{sec:6.4}, followed by evaluations under additional scenarios in Section~\ref{sec:6.5}. Due to space constraints, we defer prompt templates and additional results, including API selection, overhead analysis, and ablation of subspace formulation to  appendix~\cite{appendix}.

\subsection{Experimental Setup}\label{sec:6.1}

\textbf{Deployment Setting.}
To evaluate the generality of \sys across different database engines, we conduct experiments on two open-source database systems, PostgreSQL and TiDB.
We evaluate \sys on PostgreSQL deployed on a cloud ECS machine equipped with 16 vCPUs, 64GB of RAM, and one NVIDIA RTX 5880 Ada Generation GPU featuring 48GB of memory.
The \textit{pg\_hint\_plan}~\cite{pg_hint_plan} extension is used to enable hint-based plan tuning, and QED~\cite{QED} is adopted for query equivalence checking. The experimental setup for TiDB is described separately in Section~\ref{sec:tidb}.
Both databases are initialized using the default knob settings recommended in the official deployment manual, with only primary key indexes created and no query hints applied.

\noindent\textbf{Workloads.}
We evaluate three analytic benchmarks: (1) JOB~\cite{Job}, 113 queries based on real workloads over the IMDB dataset;
(2) TPC-H SF10~\cite{tpch}, a decision-support benchmark containing 22 analytical queries.
(3) TPC-DS SF50~\cite{tpcds}, a data-warehousing benchmark with 99 complex queries over 25 tables. These benchmarks cover diverse data scales and workload patterns.

\noindent\textbf{Tuning Space.}
We construct the knob space using 27 query-level tunable knobs (e.g., max\_parallel\_workers, work\_mem) and use 19  query hints (e.g., enable\_hashagg, enable\_sort) to form hint space. For index space, \sys considers indexes over all database columns including composite indexes.
For query rewrite, we consider 70 rewrite rules augmented with descriptions as guidance~\cite{llm4rewrite_rules}.

\noindent\textbf{Baselines.}
We compare two categories of baselines: \emph{single-component methods} and \emph{multi-component methods}, covering both specialized tuners and unified tuning frameworks. For fair comparison, all LLM-based baselines  use GPT-4.1 as the backend model.

\noindent\textit{\underline{Single-Component Baselines.}}
We include four representative tuning methods, each specialized for a single tuning component:
\begin{itemize}[leftmargin=*]
    \item \textbf{LLMIA}~\cite{LLMIdxAdvis}: an LLM-based index advisor that recommends index configurations based on workload and database features.
    
    \item \textbf{GPTuner}~\cite{GPTuner}: an LLM-enhanced knob tuning system that leverages structured knowledge extracted from DBMS manuals and related text to guide workload-aware knob selection with coarse-to-fine Bayesian tuning optimization.
    
    \item \textbf{R-Bot}~\cite{R-Bot}: an LLM-based rewriting system that retrieves rewrite evidence and iteratively refines rewriting decisions.
    
    \item \textbf{AutoSteer}~\cite{AutoSteer}: a learned query hint recommendation system that automatically discovers effective hint sets to steer the optimizer.
\end{itemize}

\noindent\textit{\underline{Multi-Component Baselines.}}
We compare \sys against existing multi-component tuning baselines:
\begin{itemize}[leftmargin=*]
    \item \textbf{Sequential Tuning}: A composite baseline that combines LLMIA, GPTuner, AutoSteer, and R-Bot to optimize all four components respectively in a  sequential order. For fairness, the total tuning budget is evenly divided across the four components.
    \item \textbf{UDO}~\cite{UDO}: A system that relies on \emph{predefined tuning schedules}. It separates heavy parameters (e.g., indexes) and light parameters (e.g., knobs), and optimizes them with different reinforcement learning strategies. It formulates tuning as a two-level process, where heavy parameters are optimized first, followed by the optimization of light parameters conditioned on the selected heavy-parameter configuration.
    
    \item \textbf{UniTune}~\cite{Unitune}: A framework that coordinates multiple tuners for different components and selects which tuner to execute based on observed feedback. 
    
    \item \textbf{Proto-X}~\cite{ProtoX}: A  system  that treats all configurable components as a unified search space and employs an actor-critic model to synthesize tuning actions in a single step. It captures structural and objective similarities across heterogeneous configuration spaces in a latent space to improve search efficiency.
    \item \textbf{Booster}~\cite{booster}: An  LLM-based framework that adopts a retrieval-augmented generation (RAG) paradigm, organizing past observations into query-level contexts and retrieving relevant historical cases for each target query based on similarity.

\end{itemize}

\noindent\textbf{Evaluation Protocol.}
We set the overall tuning budget to 30 hours and impose a 5-minute timeout for each query execution. Queries with execution latency below 1 second are considered normal and excluded from the tuning process. 
We report the average query latencies as the evaluation metric, 
with each result averaged over 3 trials.
For fair comparison, we follow Proto-X to execute the workload queries serially during evaluation.

\begin{figure*}[t]
    \centering 
    \makebox[\textwidth][c]{%
    \includegraphics{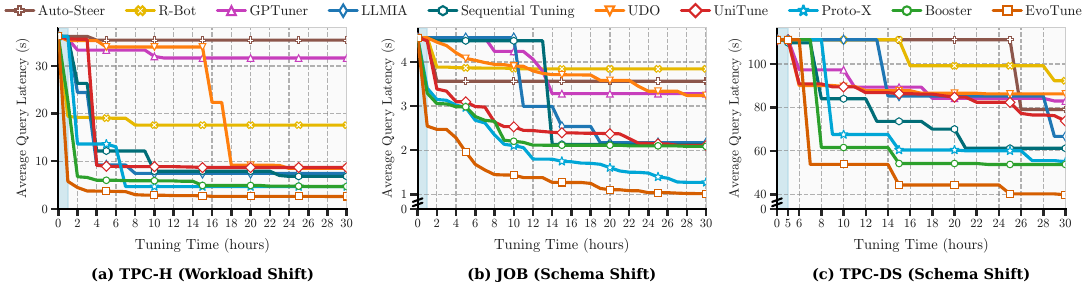}
    } 
    \caption{Best-so-far average query latency (lower is better) versus tuning time. }
    \label{fig:performance}
\end{figure*}

\noindent\textbf{\sys Setting.}
For the tuning space localization module, we adopt the model architecture of RCRank~\cite{RCRank} for the lightweight model and continuously update it with a masked BCE loss on incrementally collected data, using a batch size of 8 (Section~\ref{sec:4.1.2}).
For the tuning action recommendation module, we set $k=4$ for top-$k$ observation selection, since prior studies~\cite{few-shot} show that selecting 3--5 observations provides a practical balance between diversity and relevance. To preserve contrastive richness, we limit the selection to at most 2 positive and 2 negative observations. For the Verification Agent, we implement a maximum retry budget of 3 times.
For the learned retriever, we set the discount factor $\gamma$ to 0.99 and update the Q-network using randomly sampled mini-batches of 64 samples.
To organize the experience, we cluster observations using HDBSCAN ($\textit{min\_cluster\_size}=5$, $\textit{min\_samples}=5$), and maintain a pending buffer of size 20 to trigger periodic re-clustering.

\noindent\textbf{Offline Preparation.}
\sys collects tuning trajectories (30 hours, 3,385 samples), which are used to pre-train the component scoring model (Section~\ref{sec:4.1.1}) and the retrieve model (Section~\ref{sec:5.2}) on synthetic workloads derived from TPC-H schemas.
To evaluate whether the two models in \sys can generalize to other scenarios, we consider both workload shift and schema shift settings in Section~\ref{sec:6.2}. Specifically, we evaluate on the standard TPC-H workload (workload shift), as well as JOB and TPC-DS workloads (schema shift).
We further report a cold-start setting (Section~\ref{sec:tidb}), where neither model is pre-trained and both are updated purely online during tuning, to evaluate their continuous learning ability under more challenging scenarios.
For Proto-X, we follow its original paper~\cite{ProtoX} and construct the latent space using 8192 samples.
In contrast, the memory hub in both \sys and Booster is initialized empty to ensure a fair comparison and is progressively populated during tuning.


\begin{figure*}[t]
    \centering 
    \includegraphics{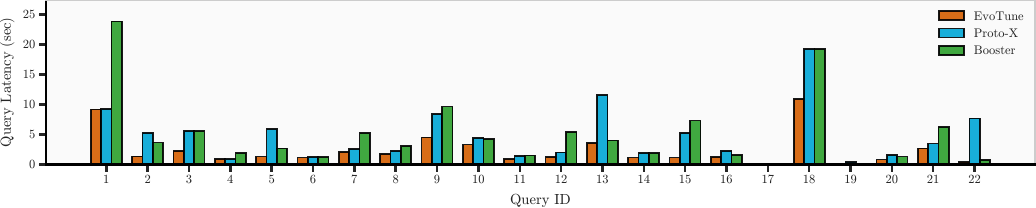} 
    \caption{Per-query performance comparison on the TPC-H workload.}
    \label{fig:performance_per_query}
\end{figure*}

\subsection{End-to-end Comparison}\label{sec:6.2}

We first report the end-to-end comparison result.
As illustrated in Figure~\ref{fig:performance}, \sys consistently outperforms all baseline approaches across all workloads  under both workload shift and schema shift settings.
Single-component baselines are effective only when the dominant bottleneck lies within their optimization scope, leading to limited performance gains. This highlights that, in realistic workloads, performance bottlenecks often span multiple components, underscoring the necessity of multi-component tuning.

\noindent\textbf{Analysis of Multi-Component Baselines.}
Sequential Tuning evenly allocates a fixed tuning budget across components, wasting efforts on ineffective subspaces and leading to suboptimal exploration when bottlenecks vary across queries.
Among multi-component baselines, UDO performs the worst due to its fixed tuning schedule, which spends a large portion of the budget exploring the knob space regardless of query characteristics, making it unsuitable for diverse workloads. UniTune shows limited performance in the early stage, especially in TPC-DS, as it relies on random subspace selection before sufficient feedback is accumulated.
Proto-X suffers from the curse of dimensionality, requiring extensive trial-and-error exploration in a large tuning space.

Despite being an LLM-based baseline, Booster performs significantly worse than Proto-X on the JOB workload. This is because it relies on high-quality historical samples to construct its memory (in its original design~\cite{booster}, such samples are generated by running an ML-based tuner for hours), but lacks the ability to effectively explore the tuning space and continuously evolve its policies. As a result, it tends to prematurely converge to configurations that are only marginally better than those stored in its memory, rather than  discovering substantially better configurations. In our setting, where all methods start from an empty memory for fairness, this tendency toward premature convergence becomes more pronounced.
In contrast, \sys incorporates evolution into both subspace exploration and action recommendation by iteratively refining the search space and dynamically retrieving high-utility experiences, enabling more effective exploration and progressively improving decisions.
Under the same 30-hour tuning budget, \sys reduces the final average query latency by 20.3\%, 44.5\%, and 27.7\% on JOB, TPC-H, and TPC-DS, respectively. Compared with the strongest baseline on each workload---Proto-X on JOB and TPC-H, and Booster on TPC-DS---\sys reaches the same final performance 2.1$\times$, 3.9$\times$, and 1.5$\times$ faster, respectively.

\noindent\textbf{Per-Query Performance Analysis.}
We  further analyze per-query execution time on the TPC-H benchmark. \sys is particularly effective at reducing the execution time of long-tail queries (e.g., Q5, Q9, and Q15). In contrast, Proto-X and Booster exhibit persistently high execution times on these complex queries, indicating their inability to resolve specific bottlenecks. By better coordinating multiple tuning components, \sys substantially improves the performance of these hard queries.



\noindent\textbf{Overhead Analysis.} 
The   tuning overhead  can be divided into two parts: (1) \textit{algorithm overhead} and (2) \textit{replay overhead}. For algorithm overhead, \sys spends 3.8 s on diagnosis, 3.7 s on tuning action recommendation, 1.0 s for lightweight model refinement and 0.6 s on online retriever update on average for each SQL. 
Although EvoTune may incur higher  algorithm overhead than ML-based tuners, it reaches the same target performance with substantially lower replay cost.
To measure the efficiency, we define a target as achieving 80\% of EvoTune’s best performance, as higher targets are often unattainable for other methods, and measure the time required to reach this target among baselines.
On TPC-H, EvoTune incurs 2.4$\times$ higher algorithm time than Proto-X to reach the target, but reduces replay time by 2.5$\times$, resulting in a 2.1$\times$ shorter total time-to-target. In comparison to Booster, EvoTune uses 3.8$\times$ less algorithm time and 1.9$\times$ less replay time, yielding a 2.3$\times$ shorter total time. Similar trends hold across the other workloads (Appendix D~\cite{appendix}).
Overall, although \sys incurs higher algorithm overhead than ML-based tuners, it is offset by a substantially lower replay cost, which dominates the total tuning time, as \sys requires far fewer query executions to reach high-quality configurations.

\subsection{Evaluation on Space Localization}\label{sec:6.3}

\begin{figure}[t]
    \begin{minipage}{0.235\textwidth}
        \centering
        \includegraphics{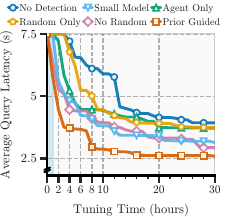}
        \caption{Detection Analysis.}
        \label{detection}
    \end{minipage}
    \hfill
    \begin{minipage}{0.235\textwidth}
        \centering
        \vspace{0.21cm}
        \includegraphics{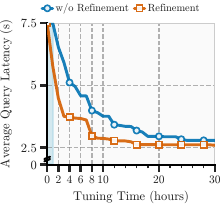}
        \caption{Refine Analysis.}
        \label{continual_a}
    \end{minipage}
    \vspace{-0.4cm}
\end{figure} 

\begin{figure}[t]
    \begin{minipage}{0.235\textwidth}
        \centering
        \vspace{4mm}
        \includegraphics{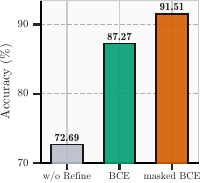}
        \vspace{0.15cm}
        \caption{Refine Variants.}
        \label{updater}
    \end{minipage}
    \hfill
    \begin{minipage}{0.235\textwidth}
        \centering
        \includegraphics{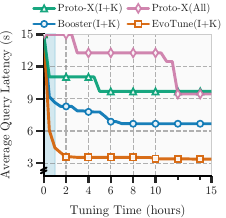}
        \caption{Tuning Analysis.}
        \label{downstream}
    \end{minipage}
\end{figure}

\subsubsection{Ablation on Collaborative Detection.}
Figure~\ref{detection} compares the
tuning performance of different variants for tuning-space localization:
(1) \textit{Prior-Guided Selection}, which adaptively chooses among the lightweight model, the Detection Agent, and random exploration;
(2) \textit{Agent Only}, which relies solely on the Detection Agent;
(3) \textit{Small Model Only}, which uses only the lightweight model; 
(4)  \textit{Random Only}, which relies solely on random exploration;
(5) \textit{No Random}, which disables random fallback;
(6) \textit{No Detection}, which directly recommends tuning actions without explicit tuning-space localization.
The results show that \textit{No Detection} suffers from sluggish convergence and easily gets trapped in local optima due to unguided exploration. 
Both \textit{Agent Only} and \textit{Small Model Only} improve over \textit{No Detection}, but each remains limited by its own inductive bias: \textit{Agent Only} captures query semantics and DBA knowledge, yet may overlook implicit execution regularities, while \textit{Small Model Only} gradually learns recurring patterns from feedback but lacks the semantic reasoning needed for robust localization. 
\textit{No Random} achieves rapid initial improvement, but soon stagnates at a local optimum because the search becomes overly exploitative. By adaptively balancing efficient pattern-based scoring, knowledge-driven reasoning, and random fallback, \textit{Prior-Guided Selection} consistently achieves both faster convergence and better final performance.


\subsubsection{Evaluation on Continual Refinement.}
We evaluate the design of continual refinement of the detection model from two perspectives: tuning effectiveness and localization accuracy. As shown in Figure~\ref{continual_a}, enabling \textit{Refinement} consistently reduces average query latency and reaches strong performance earlier than \textit{w/o Refinement}, indicating that refinement improves subspace localization throughout tuning. 
For localization accuracy, we evaluate accuracy on the target workload using ground-truth labels obtained with a sufficiently large tuning budget. 
To study the impact of different refinement objectives, we also compare \textit{BCE} and \textit{masked BCE}.
As shown in Figure~\ref{updater}, both \textit{BCE} and \textit{masked BCE} improve over \textit{w/o Refine}, and \textit{masked BCE} achieves the highest accuracy of 91.51\%. This confirms that masking unreliable negative feedback is critical for online update, as it reduces false-negative bias and yields more accurate component localization.

\subsection{Evaluation on  Action Recommendation}\label{sec:6.4}

\subsubsection{Comparison of Tuning Policies within the Same Subspace.}
Having shown the importance of subspace localization, we isolate tuning policy effects by comparing methods within the same subspace.
We evaluate tuning efficiency on TPC-H SF10 using 7 queries with index- and knob-related anomalies, comparing the top two baselines (Proto-X and Booster) within the same I+K subspace, with \textit{Proto-X(All)} as a full-space baseline.
As shown in Figure~\ref{downstream}, restricting Proto-X to the I+K subspace improves convergence over \textit{Proto-X(All)}, highlighting the benefit of search space reduction.
\textit{EvoTune} achieves the best performance by more effectively exploiting observations and experiences, reaching strong configurations within the first few tuning rounds.

\begin{figure}[t]

    \begin{minipage}{0.235\textwidth}
        \centering
        \includegraphics{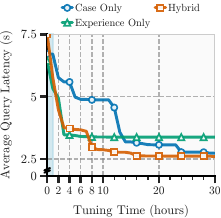}
        \caption{Memory  Effect.}
        \label{retrieve_level}
    \end{minipage}
    \hfill
    \begin{minipage}{0.235\textwidth}
        \centering
        \includegraphics{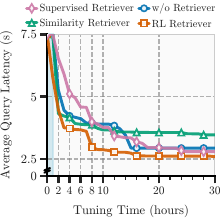}
        \caption{Retriever Analysis.}
        \label{retrieve_effect}
    \end{minipage}
\end{figure} 

\begin{figure}[t]
    \begin{minipage}{0.235\textwidth}
        \centering
        \vspace{2mm}
        \includegraphics[width=0.9\textwidth]{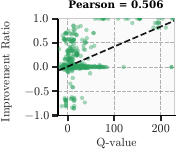}
        \vspace{1mm}
        \caption{RL Retrieve Scatter.}
        \label{retrieve_scatter}
    \end{minipage}
    \hfill
    \begin{minipage}{0.235\textwidth}
        \centering
        \includegraphics[width=0.9\textwidth]{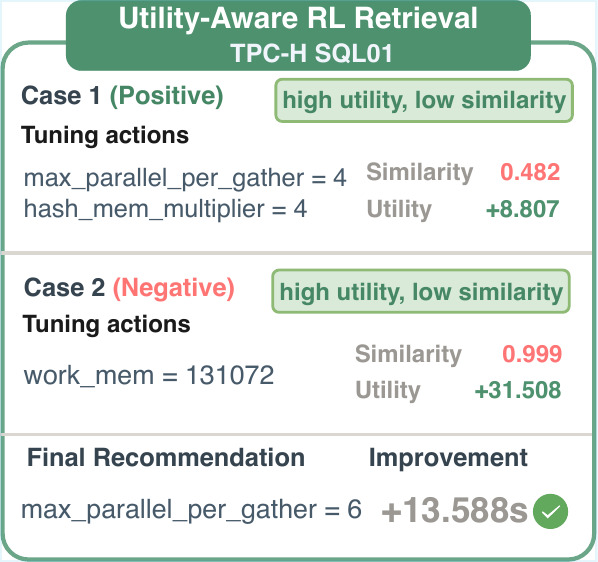}
        \caption{Retrieval Example.}
        \label{retrieve_example}
    \end{minipage}
\end{figure}

\subsubsection{Evaluation on Memory Construction}
\sys constructs a hybrid memory hub that integrates both fine-grained cases and generalized experience.
To validate this design, we compare three variants: (1) \textit{Case Only}, which uses only fine-grained cases; (2) \textit{Experience Only}, which relies solely on generalized experience; and (3) \textit{Hybrid}, our design that integrates both.
As shown in Figure~\ref{retrieve_level}, the hybrid memory outperforms both \textit{Experience Only} and \textit{Case Only}, as it combines high-level guidance from coarse-grained experiences with fine-grained signals from individual cases.

\subsubsection{Retrieval policy comparison.}
We compare four variants of the retriever module:
(1) \textit{w/o Retriever}, which performs zero-shot tuning without leveraging historical observations;
(2) \textit{Similarity Retriever}, which retrieves observations based on cosine similarity between queries;
(3) \textit{Supervised Retriever}, which ranks observations by predicting their utility given the current tuning context; and
(4) \textit{RL Retriever}, our approach that learns a utility-aware policy to prioritize observations with higher long-term impact.
As shown in Figure~\ref{retrieve_effect}, removing the retriever forces the Recommendation Agent to rely on unguided search, leading to inferior performance.
The \textit{Similarity Retriever} partially alleviates the cold-start issue. 
However, as tuning progresses, it repeatedly retrieves the same observations that provide limited or even misleading utility, causing the similarity-based approach to converge prematurely.
The \textit{Supervised Retriever} further improves by learning from data, but it optimizes for immediate utility and does not account for long-term effects through state transitions, limiting its ability to capture sequential dependencies.
In contrast, the \textit{RL Retriever} learns to prioritize observations with higher downstream utility. As illustrated in Figure~\ref{retrieve_scatter}, the learned Q-values exhibit a stronger correlation with the final improvement ratio than similarity (Figure~\ref{fig:challenge_c}). Although the overall correlation is moderate---particularly for low-Q regions---the observations with higher Q-values are consistently associated with positive performance gains, indicating that the retriever can effectively identify high-utility observations.
Figure~\ref{retrieve_example} further provides a concrete example: the RL retriever recalls a high-utility case despite low cosine similarity, enabling the agent to synthesize a better recommendation and achieve substantial improvement.




\subsection{More Scenarios}
\label{sec:6.5}




\begin{figure}[t]
    \begin{minipage}{0.235\textwidth}
        \centering
        \vspace{1mm}
        \includegraphics{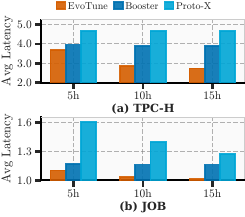}
        \vspace{-5mm}
        \caption{Initialization from Existing Tuner.}
        \label{scenario1}
    \end{minipage}
    \hfill
    \begin{minipage}{0.235\textwidth}
        \centering
        \includegraphics{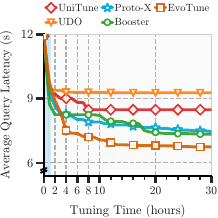}
        \caption{Cold Start on TiDB.}
        \label{tidb}
    \end{minipage}
\end{figure}


\subsubsection{Initialization from Existing Tuner.}
We further evaluate a warm-start setting where tuning begins from a pre-populated memory derived from an existing tuner. This setting is motivated by Booster, which is designed to assist existing tuners by leveraging pre-collected observations.
To ensure a fair comparison, we initialize all methods from a memory hub constructed by running Proto-X for 12 hours, following Booster’s original setup. Figure~\ref{scenario1} reports the results.
Recall that when starting from an empty memory (Section~\ref{sec:6.2}), Booster performs worse than Proto-X on JOB workload due to its limited exploration capability. In contrast, under this warm-start setting, Booster benefits from the initial exploration provided by Proto-X and outperforms it.
EvoTune consistently achieves the best performance, benefiting from its ability to continuously evolve tuning policies through utility-aware retrieval and memory updates. 

\subsubsection{Robustness of Evolution under Cold Start (TiDB)}\label{sec:tidb}
To evaluate whether \sys can bootstrap memory-aware evolution on a previously unseen engine, we consider a cold-start setting where both the detection model and the retriever are initialized without offline pre-training, and the memory hub is empty at the start.
We conduct experiments on TPC-H SF10 using a TiDB v8.5.3 cluster with 1 TiDB, 1 PD, 1 TiFlash, and 3 TiKV nodes. The TiDB tuning space includes 19 query-level knobs and 11 optimizer hints, while the index and query rewrite spaces follow the initialization strategy in Section~\ref{sec:6.1}.
As shown in Figure~\ref{tidb}, \sys underperforms in the first three hours due to the cold-start setting, where both the detection model and the retriever are initialized without prior training, limiting retrieval quality and localization accuracy.
As tuning proceeds, newly collected observations are written back to memory and reused, enabling improved localization and more effective optimization.
 Consequently, \sys surpasses all baselines from the fourth hour onward. These results show that \sys can bootstrap effective memory online and steadily improve even without  offline engine-specific experience.

\section{Related Work}\label{sec:2.3}

\noindent\textbf{Anomaly Detection.}
Database anomaly detection seeks to identify the root causes of query performance degradation. 
Data-driven methods~\cite{DBSherlock,Dbmind,iSQUAD,RCRank} use system metrics, query plans, or execution statistics for diagnosis, but often generalize poorly under workload shifts due to their reliance on labeled data. 
Knowledge-driven methods~\cite{Panda,DBG-PT,D-Bot} leverage LLMs to improve diagnostic reasoning, yet mainly depend on static prior knowledge and fail to utilize feedback signals for continual refinement. 
Besides, some single-component tuning approaches reduces the search space within individual components (e.g., knob selection~\cite{DBLP:conf/hotstorage/KanellisAV20}) but fails to capture component importance and interactions.
In contrast, we identify anomaly-related components to localize a query-specific anomaly subspace.


\noindent\textbf{LLM-Empowered DBMS Tuning.}
Recent studies have explored LLM-based agents for DBMS tuning across different tasks, including knob tuning~\cite{GPTuner, E2ETune, Andromeda, AgentTune}, query rewriting~\cite{R-Bot, LLM-R2}, and index tuning~\cite{LLMIdxAdvis}. However, these approaches typically focus on \emph{single-component} optimization. 
Methods such as~\cite{GPTuner, Andromeda, R-Bot} encode manual knowledge into prompts, providing useful but largely static guidance.
Other approaches~\cite{E2ETune,AgentTune,LLMIdxAdvis,LLM-R2} incorporate historical feedback, but exploit it only in limited ways. Specifically, E2ETune~\cite{E2ETune}  improves tuning ability via offline fine-tuning on historical observations, but incurs high data preparation cost and generalizes poorly under workload shifts. AgentTune~\cite{AgentTune} reduces this overhead by leveraging online feedback, but considers only top-$k$ observations for the current query. LLMIdxAdvis~\cite{LLMIdxAdvis} and LLM-R2~\cite{LLM-R2} rely on static similarity-based retrieval, which surface low-utility cases and lack mechanisms for continual adaptation.
To the best of our knowledge, Booster~\cite{booster} is the only LLM-based method that considers \emph{multi-component} tuning. It retrieves query-level historical contexts and suggests per-query recommendations. 
However, it explores the entire configuration space without effective localization and still relies on similarity-based retrieval, making it prone to inefficient exploration and low-utility guidance.
\section{Conclusion}
Existing multi-component DBMS tuning approaches often suffer from unguided exploration and high iterative overhead.
To address these limitations, we propose \sys, an LLM-empowered framework for autonomous multi-component DBMS tuning.
\sys localizes high-impact tuning subspaces via anomaly detection and refines tuning policies through utility-aware retrieval.
Moreover, it introduces a memory-aware mechanism that continuously exploits accumulated experience to jointly improve anomaly diagnosis and tuning decisions without fine-tuning the LLMs.
Extensive experiments show that \sys consistently outperforms state-of-the-art baselines, achieving up to 44.5\% performance improvement under the same tuning budget while reaching the best competing baseline performance with up to 3.9$\times$ less tuning time.
\balance
\bibliographystyle{ACM-Reference-Format}
\bibliography{references}

\end{document}